\documentclass[tcfd]{svjour}
\usepackage{times}
\usepackage{graphics}
\begin{document}

\title{Complex dynamics in double-diffusive convection}
%
\Author{Esteban Meca}{Departament de F\'{\i}sica Aplicada, Universitat
Polit\`ecnica de Catalunya, Doctor Mara\~n\'on 44, E-08028 Barcelona, Spain}
\Author{Isabel Mercader}{Departament de F\'{\i}sica Aplicada, Universitat
Polit\`ecnica de Catalunya, Jordi Girona 1-3, E-08034 Barcelona, Spain}
\Author{Oriol Batiste}{Departament de F\'{\i}sica Aplicada, Universitat
Polit\`ecnica de Catalunya, Jordi Girona 1-3, E-08034 Barcelona, Spain}
\Author{Laureano Ram\'{\i}rez-Piscina}{Departament de F\'{\i}sica Aplicada,
Universitat
Polit\`ecnica de Catalunya, Doctor Mara\~n\'on 44, E-08028 Barcelona, Spain}
\commun{Communicated by }
%
\date{Received date and accepted date}
%

 \abstract{ The dynamics of a small Prandtl
number binary mixture in a laterally heated cavity is studied
numerically. By combining temporal integration, steady state solving
and linear stability analysis of the full PDE equations, we have
been able to locate and characterize a codimension-three
degenerate Takens-Bogdanov point
 whose unfolding describes the
dynamics of the system for a certain range of Rayleigh numbers and
separation ratios near S=-1.}

\authorrunning{Esteban Meca et al.}
\maketitle
%


\section{Introduction}
\label{sec:intro} Double-diffusive fluxes occur when convection is
driven by thermal and concentration gradients, and the temperature
and concentration diffusivities take different values.  This
phenomenon has relevance for numerous applications
\citep{Turner85}, and from a theoretical point of view presents
very interesting dynamics, including chaos
\citep{cross93,KnMoToWe86}.  We are interested here in the case of
horizontal gradients \citep{Turner80,Jiang91}. In this configuration, quiescent
(conductive) solutions can exist when thermal and solutal buoyancy
forces exactly compensate each other. This occurs only for a very
particular value of the separation ratio ($S=-1$, see below), but
has allowed for theoretical analysis by studying the stability of
the conductive solution \citep{GhMo97,XiQeTu97,BaBeKnMo00,BeKn02}.
In recent work, we addressed this case for a small Prandtl number
binary mixture, including only the Soret effect \citep{mecaprl}.
Results showed a quite interesting bifurcation scenario by varying
the Rayleigh number. In particular, we found an orbit that is born
in a global saddle-loop bifurcation, becomes chaotic in a period
doubling cascade, and disappears in a blue sky catastrophe
\citep{LShil97}. This orbit is the only stable solution in a large
interval of Rayleigh numbers. In this paper we analyze this system
in greater depth to determine how the scenario associated to the
origin of this orbit is modified when the value of $S$ is changed
to larger ({\it i.e.} less negative) values. It is relevant to
assess to what extent the dynamics depends on tuned values of
the parameters, or whether it is fairly robust to these
changes.  Moreover, by varying a second parameter we gain access to
a richer portrait of the system, obtaining bifurcation lines and
points of codimension two.

 We have numerically integrated the full PDE equations in a
region near $S=-1$, combining
steady state solving, numerical continuation,
linear stability analysis,  and temporal integration.
The results show that distinct bifurcations
of the $S=-1$ case (namely two saddle nodes and a global saddle loop)
approach each other in the region near $S=-0.9$ until only a Hopf
bifurcation remains, in a scenario consistent with the unfolding of a
codimension-three degenerate Takens-Bogdanov point.

The outline of this paper is as follows. In Section \ref{sec:eqs} we
detail the model and the numerical procedure. In Section
\ref{sec:seq-1} the behavior of the system for $S = -1$ is
reviewed. In Section \ref{sec:sgt-1} we extend these results by
letting both $Ra$ and $S$ vary. Finally, the discussion of the results
and some concluding remarks are presented in Section \ref{sec:concl}.


\section{Basic equations and numerical methods}
\label{sec:eqs}

We consider the 2-D flux of a binary mixture in a rectangular
cavity $\Omega$ of length $d$ and height $h$. The aspect ratio $\Gamma = d/h$
has been chosen to be $2$. The cavity is laterally heated, maintaining
different constant temperatures at the opposed vertical
boundaries. $\Delta T$ is the difference between both temperatures. On
the horizontal boundaries, a linear temperature profile is imposed. All
the boundaries are taken to be no-slip and with no mass flux. In these
conditions the dimensionless equations in Boussinesq approximation
explicitly read
\begin{eqnarray}
 \partial_t {\bf u} + ({\bf u} \cdot \nabla){\bf u}  &  =
& -\nabla P  + \sigma \nabla^2{\bf u}
\nonumber
\\
 & + & \sigma Ra [
\left(1+S\right)\left(-0.5 +x/\Gamma\right) + \theta + S C ] \hat{\bf z},
\nonumber
\\
 \label{ecs}
 \partial_t \theta + ({\bf u} \cdot \nabla) \theta  & =
& - v_x/ \Gamma + \nabla^2 \theta,
\\
 \partial_t C + ({\bf u} \cdot \nabla) C  & =
& - v_x /\Gamma - \tau \nabla^2
(\theta - C),
\nonumber
\\
 \nabla \cdot {\bf u}  & =
& 0.
\nonumber
\end{eqnarray}
In these equations lengths, times and temperatures are scaled with $h$,
$t_{\kappa}=h^2/\kappa$ and $\Delta T$, respectively, $\kappa$ being the thermal
diffusivity. ${\bf u} \equiv (v_x,v_z)$ is the (dimensionless)
velocity field in $(x,z)$ coordinates, $P$ is the pressure over the
density,
$\theta$ and $C$ are deviations from a linear horizontal profile of the temperature and of
the rescaled concentration of the heavier component, respectively.
The dimensionless
parameters are the Prandtl number ${\sigma}=\nu / \kappa$, the
Rayleigh number ${ Ra}={\alpha g h^3} \Delta T / {\nu \kappa}$ and the
Lewis number ${\tau} = {D} / {\kappa}$, where $\nu$ denotes the
kinematic viscosity, $g$ the gravity level, $\alpha$ the thermal
expansion coefficient, and $D$ is the mass diffusivity.  The
separation ratio is defined by $S= C_0(1-C_0)\frac{\beta}{\alpha}S_T$,
where $S_T$ is the Soret coefficient, $C_0$ is the actual value of the concentration of the
heavier component in the homogeneous mixture, and $\beta$ is the
mass expansion coefficient (positive for the heavier
component). Finally, boundary conditions are written as
\begin{equation}\label{bc}
{\bf u}={\theta}= {\bf n} \cdot \nabla (C - \theta)=0, \quad \mbox
{at $\partial \Omega$}.
\end{equation}
These boundary conditions are not compatible with the transformation of the
Soret equations into those used by \citet{GhMo97,XiQeTu97,BaBeKnMo00,BeKn02}.
Note also that
Eqs. (\ref{ecs}) together with boundary conditions (\ref{bc}), are
invariant under a rotation $\pi$ around the point $(\Gamma/2,1/2)$
as
$
(x,z) \rightarrow (\Gamma -x,1-z),
(v_x,v_z,\theta,C) \rightarrow  (-v_x,-v_z,-\theta,-C) $.
Therefore the system is ${ Z}_2$-equivariant \citep{Kuznetsov}.
From now on solutions invariant (non-invariant) under $\pi$ will
be called symmetric (non-symmetric).

Eqs. (\ref{ecs}) and boundary conditions (\ref{bc}) have been
solved by a second order time-splitting algorithm, proposed by
\citet{HuRa98}, applied to a pseudo-spectral Chebyshev method.  To
calculate steady solutions, we have adapted a pseudoespectral
first-order time-stepping formulation to carry out Newton's
method, as described by \citet{MaTu95,BeHeBeTu98,XiQe01}. In the
preconditioned version of Newton's iteration, the
corresponding linear system is solved by an iterative technique
using a GMRES package \citep{cerfacs}. The linear stability
analysis of the steady states is conducted by computing the
leading eigenvalues of the Jacobian by means of Arnoldi's
method, using routines from the ARPACK package. For numerical
calculations the chosen parameters have been Prandtl number
$\sigma=0.00715$ and Lewis number $\tau=0.03$. The system has been
discretized in space by using $72 \times 48$ and $90 \times 60$
mesh points in steady calculations, giving both resolutions
equivalent results.
For example, increasing the resolution, the Rayleigh number of the turning points
varies less than a $0.1\%$
Temporal integration was used basically to
follow orbits with very long periods, and in particular in regimes
with divergent periods where a fit of the divergence itself was
needed. Thus, much more computation time was required. A
mesh grid of $60 \times 30$ points has proved to be sufficient to
obtain results that did agree fairly well with steady calculations
in the cases where the comparison applied.



\section{Scenario for $S=-1$}
\label{sec:seq-1}

In this section we summarize the behavior of the system for a
separation ratio $S=-1$ \citep{mecaprl}.
\begin{figure}
\begin{center}
\resizebox{0.75\textwidth}{!}{%
  \includegraphics{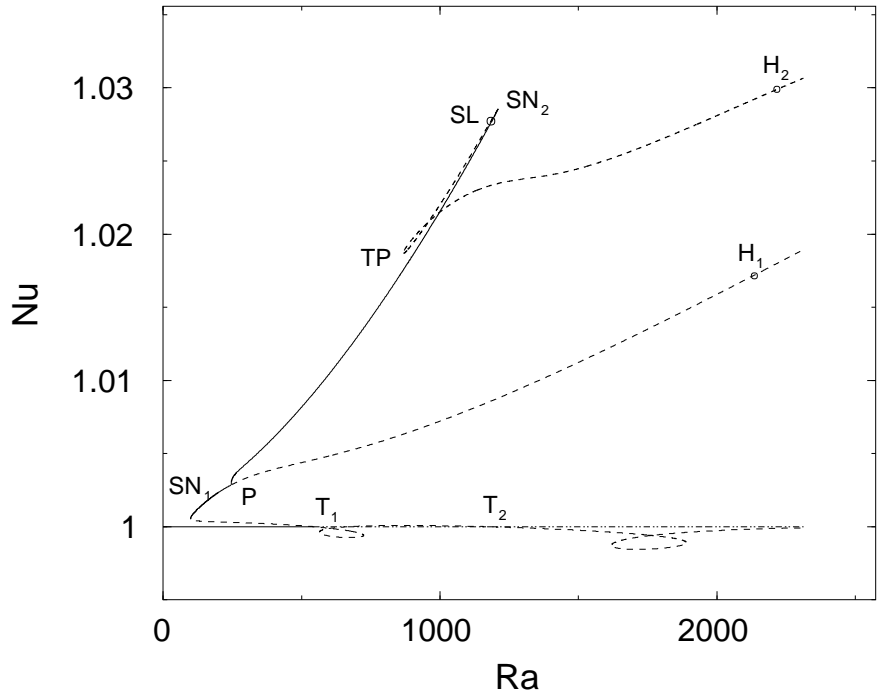}
} \caption{Stationary solutions diagram for $S=-1$ varying $Ra$,
as obtained by \citet{mecaprl}. Continuous lines: stable states.
dashed lines: unstable states. }
\label{fig.fig1}
\end{center}
\end{figure}
The bifurcations diagram of the steady solutions is shown in Fig.
\ref{fig.fig1}. In this figure, the Nusselt number $Nu$, defined
as the ratio of the heat flux through the hot wall to that of the
corresponding conductive solution, is represented as a function of
the Rayleigh number $Ra$. For the sake of clarity, only one of the
non-symmetric solutions related by $\pi$ is represented. For this
value of $S$ the conductive solution is allowed. For small $Ra$
this solution is stable, but loses stability at $Ra = 541.9$
through a transcritical bifurcation ($T_{1}$). The resulting
solutions are symmetric by the rotation $\pi$, and
are characterized by a central main roll accompanied by secondary ones
in opposed corners.  The supercritical branch is stable only until
$Ra=542.4$, where a small non-symmetric branch connects it to the
conductive solutions through pitchfork bifurcations, a case analogous to
that reported by \citet{BaBeKnMo00}. We center the discussion
here on the
solutions originating from the subcritical branch. This solution is
stabilized by a saddle node bifurcation at $Ra=99$ ($SN_{1}$) and
loses stability again in a Pitchfork bifurcation at $Ra=245$ ($P$),
where a stable non-symmetric branch appears. In Fig. \ref{fig:s-1}
(left) we see that the breaking of the symmetry confines the main roll
to one lateral side.
\begin{figure}
\begin{center}
\resizebox{0.45\textwidth}{!}{%
  \includegraphics{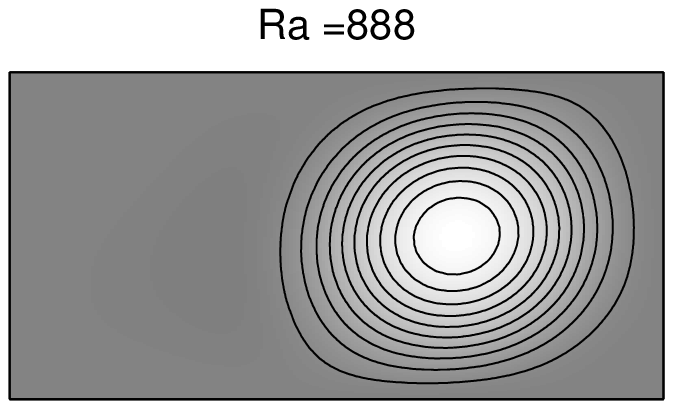}
}
\resizebox{0.45\textwidth}{!}{%
             \includegraphics{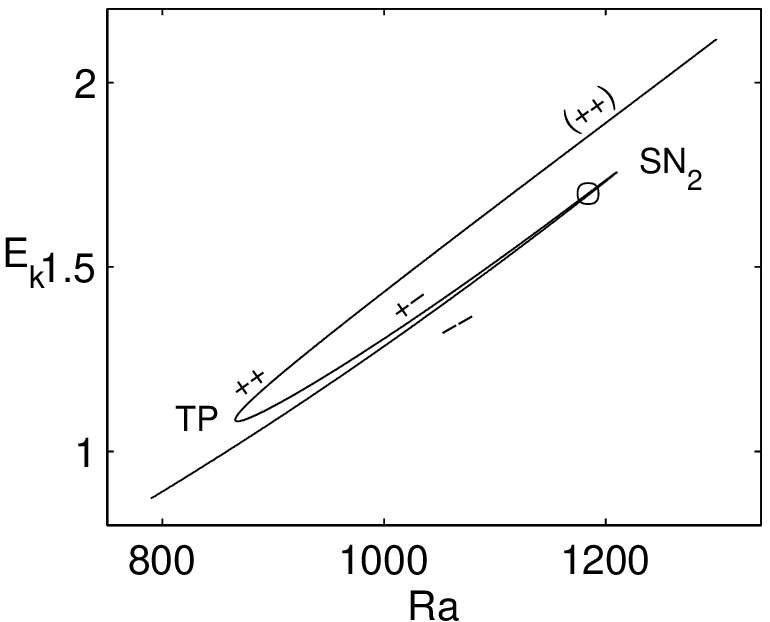}
}
\caption{Non symmetric steady solution for $S=-1$.  Left: Stream lines
for the stable solution at $Ra=888$. Right: Detail of the
non-symmetric branch in the region of the two turning points, indicating the
sign of the real part of the leading eigenvalues. The open circle denotes
a global saddle-loop bifurcation.}
\label{fig:s-1}
\end{center}
\end{figure}
%
%
Continuing the symmetric branch, a supercritical Hopf bifurcation at $Ra=2137$
($H_{1}$) (maintaining symmetry) is found.
At $Ra=2253$ the bifurcating periodic solution gains stability in a
Pitchfork bifurcation.

Furthermore, the steady non-symmetrical solution undergoes a
saddle-node bifurcation at $Ra=1209$ ($SN_{2}$) and has a saddle-saddle turning
point at $Ra=865.6$ ($TP$).  Further along this
branch, a Hopf bifurcation can be found at $Ra=2218$ ($H_{2}$).  A
detail of this branch in the region of the two turning points is
represented in Fig. \ref{fig:s-1} (right). Here, the variable $E_k$,
related with the kinetic energy and defined as
\begin{equation}
E_k= \frac{1}{\Gamma} \int_{x=0}^{x=\Gamma} \int_{z=0}^{z=1} {\bf v}{\cdot}
{\bf v}\, dx dz,
\end{equation}
 is plotted versus the Rayleigh number $Ra$.  In this figure we have
also included the sign of the real part of the leading
eigenvalues; a parenthesis is used to indicate a complex
conjugated pair. Between both turning points an homoclinic saddle
loop connection at $Ra=1184$ ($SL$) gives birth to a periodic
non-symmetric solution \citep{mecaprl}. This orbit, which is the
only stable solution in a wide range of values of the Rayleigh
number $Ra$, is characterized by very long periods and a spiking
behavior. In particular at the $SL$ connection its period diverges
logarithmically as expected.

When the Rayleigh number is increased, a very interesting complex behavior
of this orbit
arises \citep{mecaprl}. Firstly, at $Ra=2137$ the orbit starts to show ripples,
reflecting the frequency corresponding to the Hopf bifurcation $H_1$,
while the period increases dramatically. In the region near $Ra=2235$
the orbit undergoes a period doubling cascade, becoming chaotic. At
$Ra=2257.5$ the chaotic attractor disappears in a blue sky
catastrophe, in a scenario similar to that proposed by
\citet{LShilTur00} in which both length and period of an orbit diverge
at the bifurcation point.


\section{Results for $S > -1$}
\label{sec:sgt-1}

We have performed both steady state and temporally dependent
calculations of the system for different values of $S$ above $-1$.  We
have centered our research on the solutions from which the
non-symmetric orbit is born in a global saddle loop (SL)
connection. Namely, we are referring to the non-symmetric branch with
two successive turning points ($SN_2$ and $TP$), between
which the SL connection is found. This is the situation represented in
Fig. \ref{fig:s-1} (right) for $S=-1$.

\begin{figure}
\begin{center}
\resizebox{0.75\textwidth}{!}{%
  \includegraphics{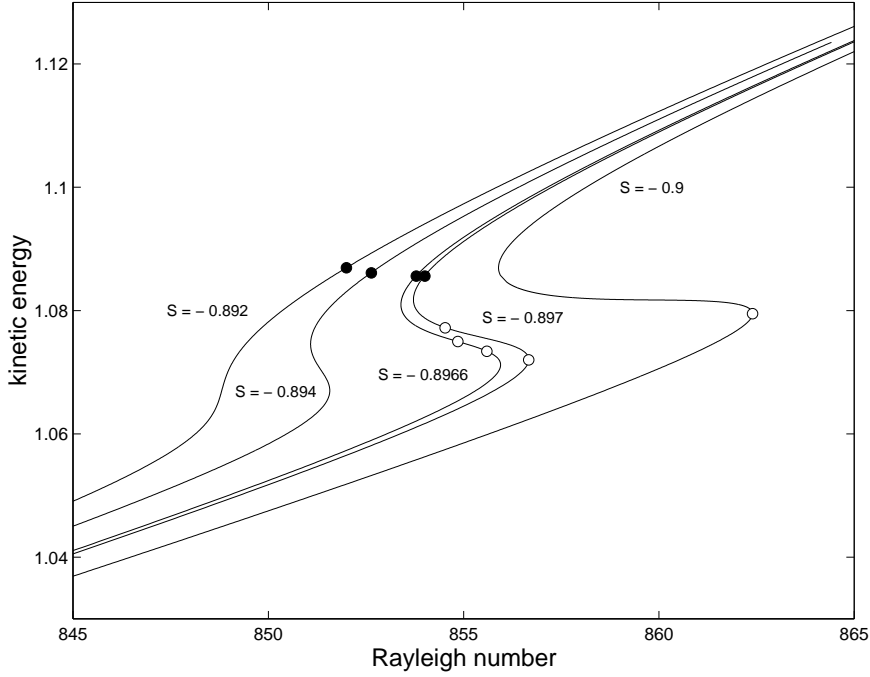}
}
\caption{Detail of the bifurcations diagram for several values
of $S>-1$. Kinetic energy is represented versus $Ra$. Full circles
denote Hopf bifurcations. Global bifurcations (either SL or SNIC)
are represented by open circles.}
\label{fig:sneq-1}
\end{center}
\end{figure}

This configuration of bifurcations changes when $S$ is increased.
The sequence of events in this region is quite complex, but the final
situation is simple.  Remarkably, for values around $S=-0.8920$ only a
Hopf bifurcation remains.  To analyze this process we present in
Fig. \ref{fig:sneq-1} results for this branch and different $S$
values.
In this figure, branches and local bifurcations are found by continuing
steady solutions and performing linear stability analysis, whereas
global bifurcations are located by temporal integration. To do this,
we fix a value of S, and starting from a periodic solution as initial
condition, we vary slightly the Rayleigh number while monitoring
the value of
the period of the final stable orbit. Then, the connection is located
at the point where the period diverges.

The first qualitative change occurs between $S=-1$ and $-0.9$,
where the global connection SL moves toward the saddle node
$SN_2$, becoming a SNIC (saddle node on an invariant circle) in a
saddle node loop (SNL) codimension-two bifurcation. This change is
manifested in the law of divergence of the period $T$ found by
temporal integration of the bifurcating orbit for $Ra$ toward the
homoclinic connection. The divergence for the SL is logarithmic,
${\it i.e.}$
\begin{equation}
T\sim-\frac{1}{\lambda}\log\left(Ra-Ra_{c}\right)+A,
\label{eq:log}
\end{equation}
where $Ra_c$ is the value for which the global connection takes
place and $\lambda$ is the eigenvalue of the jacobian matrix associated to the unstable
direction of the hyperbolic solution. The divergence changes to
square root for the SNIC:
\begin{equation}
T\sim\frac{B}{\sqrt{Ra-Ra_{c}}}+A.
\label{eq:sqrt}
\end{equation}
Here, $Ra_c$ corresponds to the position of the saddle node. Furthermore,
as we increase $S$, the point at which the two positive
eigenvalues merge to become a pair of conjugated complex values
(see Fig. \ref{fig:s-1} right) approaches the turning point ($TP$)
until a Takens-Bogdanov (TB) codimension-two bifurcation occurs
there for $S$ near $-0.8990$. In Fig \ref{fig:sneq-1} we see the
branch for $S=-0.897$, for which the $TB$ has unfolded to a Hopf
bifurcation and a global saddle loop connection, which constitute
the birth and the end of an orbit that has been calculated by
temporal integration. The divergence of the period of this orbit
at the saddle loop is again logarithmic, as in Eq. \ref{eq:log}.
 Therefore for this value of $S$ we obtain both
divergences, logarithmic for the SL and square root for the SNIC.
We have performed fits of the periods for both points. Indeed, the
$\lambda$ value obtained by the fit, $\lambda_{fit}=0.0173$,
agrees fairly well with the eigenvalue obtained by the stability
calculation of the steady state ($\lambda = 0.0166$). For the SNIC,
$Ra_c (fit) = 856.64$, and the value of $SN_2$ is $Ra_c=856.67$. In
this situation the eigenvalue that became positive at $SN_2$ gains
stability at $TP$. At the Hopf bifurcation the solution loses
stability again.

\begin{figure}
\begin{center}
\resizebox{0.45\textwidth}{!}{%
  \includegraphics{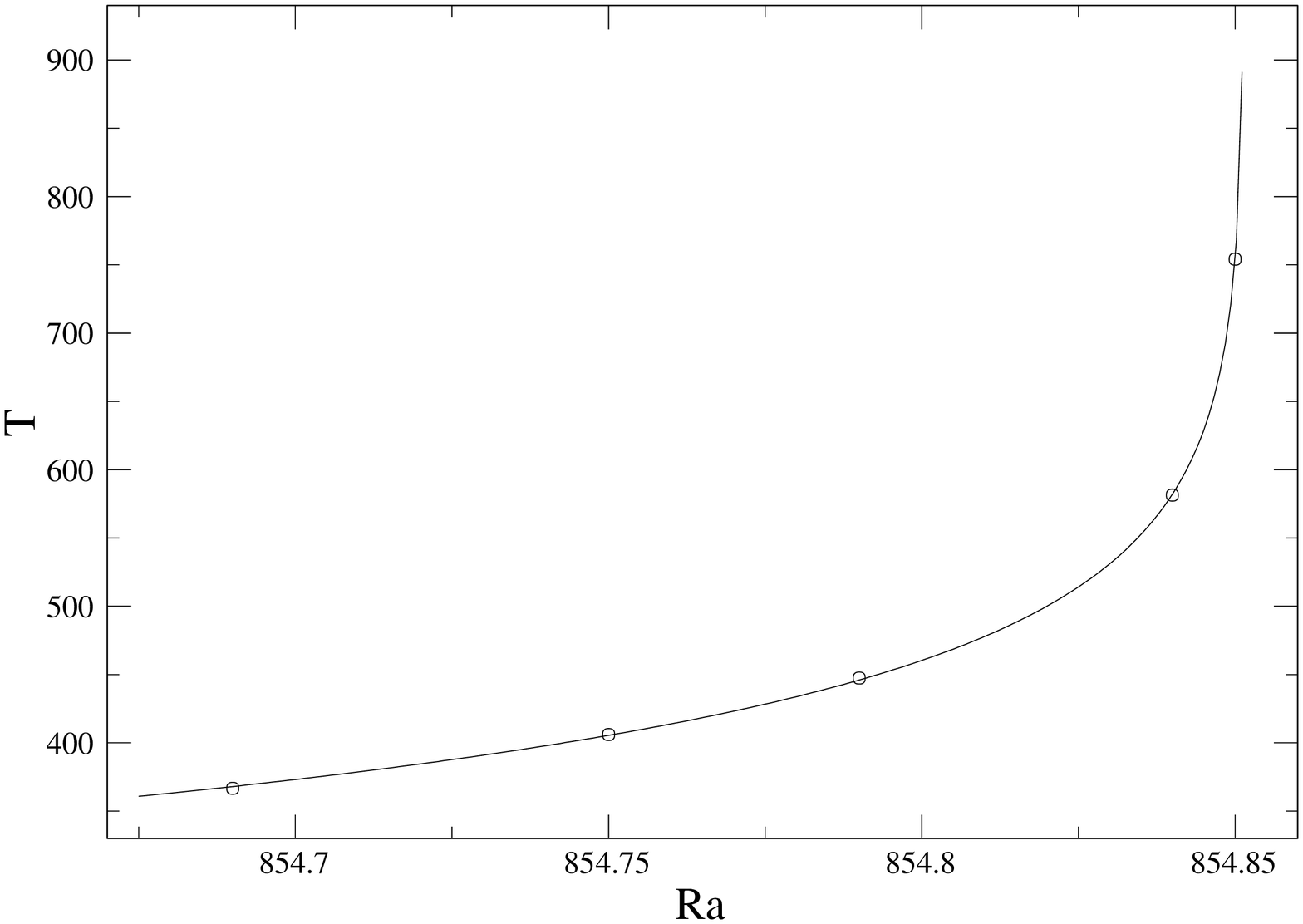}
}
\resizebox{0.45\textwidth}{!}{%
             \includegraphics{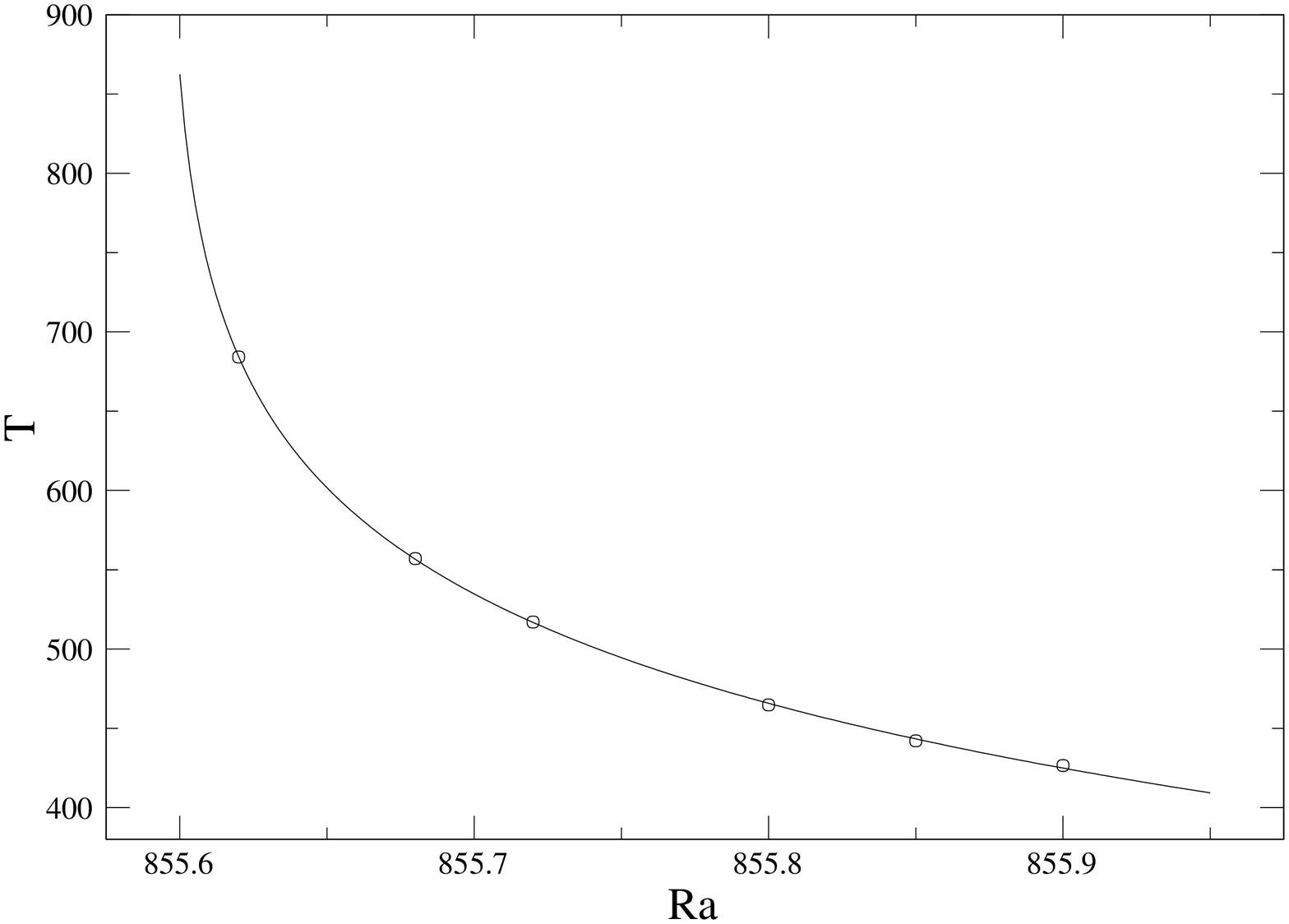}
}
\caption{Periods of the homoclinic orbits found at $S=-0.8966$
near their corresponding SL, together with the corresponding
logarithmic fits. Left: divergence located at $Ra=854.853$;
Right: divergence located at $Ra=855.595$}
\label{fig:diverg}
\end{center}
\end{figure}
When $S$ is slightly increased, an additional codimension-two SNL
bifurcation is found at $SN_2$, the SNIC becoming an SL moving away
from the saddle node. At this moment, both SL  are approaching each
other, a situation represented in the branch $S=-0.8966$ in Fig.
\ref{fig:sneq-1}. The divergences of the periods of the
corresponding homoclinic orbits are shown in Fig.
\ref{fig:diverg}.
 They are both logarithmic, with
fitted values for the eigenvalues being $\lambda_{fit}=0.00968$ and
$0.0118$, according well with the steady calculations
$\lambda=0.00820$ and $0.0134$, respectively.

 Very soon afterwards they touch each other and
disappear. By $S=-0.894$ no global connection remains in this
branch. It is now possible for the two turning points ($SN_2$ and $TP$) to annihilate
each other in a codimension-two cusp bifurcation. That occurs for
$S=-0.8928$. The final situation, in which only the Hopf
bifurcation is found, is represented by the $S=-0.892$ branch of
Fig. \ref{fig:sneq-1}.


\section{Discussion and concluding remarks}
\label{sec:concl}

In the preceding section we have numerically studied the change of
an interesting branch of non-symmetric solutions when a second
parameter (the separation ratio $S$) is varied. The scenario that
emerges from these results consists of a series of codimension-two
bifurcations arranged to enable two destabilizing saddle
nodes of a branch connected to an orbit by a global bifurcation
to disappear, resulting in a simpler situation with only a
local Hopf bifurcation.

\begin{figure}
\begin{center}
\resizebox{0.75\textwidth}{!}{%
  \includegraphics{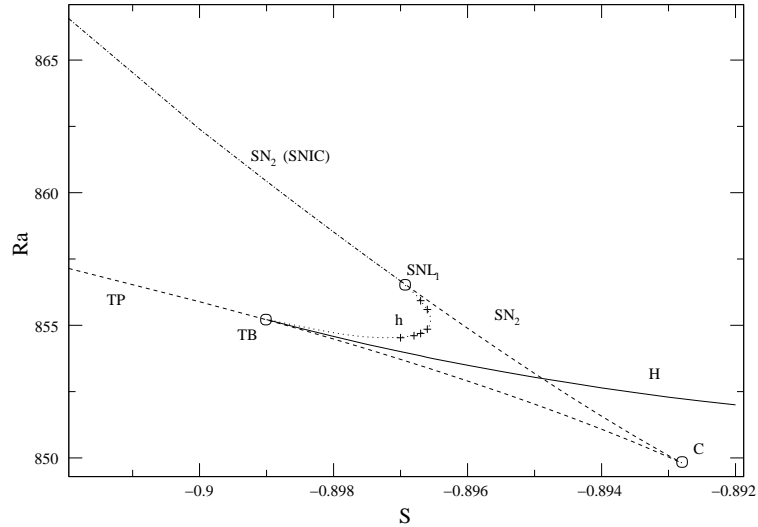}
} \caption{Bifurcation lines in the $(S,Ra)$ parameter space (see text).}
\label{fig:peix}
\end{center}
\end{figure}
These results can be summarized by drawing bifurcation lines
and codimension-two points in an $(S,Ra)$ plot. This is shown in
Fig. \ref{fig:peix}.  In this figure the two turning point lines meet in
a cusp bifurcation at $S=-0.8928$.  We can track the larger
eigenvalues of the system along these lines. We find that at
$S=-0.8990$ the solution on the line $TP$ has a double zero
eigenvalue. This indicates the presence of a Takens-Bogdanov (TB)
codimension-two point. Indeed, this can also be checked by noting that, by
performing linear stability analysis, a branch of Hopf
bifurcations, with very low frequency starting from zero, also begins
at this point.  It should be stressed that in order for both turning
points to be effectively annihilated $TP$ should become
stabilizing. That happens at the TB point.

The existence of a TB bifurcation also implies the existence of a
branch of homoclinic connections (saddle-loop). This is, in fact, one
of the few analytical methods of proving the existence of
a homoclinic orbit.
  We have used the time evolution
code to situate the line of homoclinic connections as detailed in
Sec. \ref{sec:sgt-1}, and checked if the period of the orbit
diverges following the logarithmic law (Eq. \ref{eq:log}). Indeed
the values found for $\lambda_{fit}$ did agree well with the
eigenvalues calculated from the stability of the steady solutions.
The fitting of the value of $Ra_{c}$ yields the points
marked in Fig. \ref{fig:peix}. We see that the line that
joins these points is connected with the saddle-node
line corresponding to the $SN_{2}$ points. This connection is
$SNL_1$ (located very close to $S=-0.897$), one of the saddle-node
separatrix-loop points \citep{Schecter87} mentioned above. It is
at this point that the branch of homoclinic connections merges
with the saddle-node line to become an SNIC line. This is confirmed
by the change of the logarithmic law for the divergence of the
period (Eq. \ref{eq:log}) to the square root of Eq. \ref{eq:sqrt}.

If we keep following the SNIC line toward more negative separation
ratios, we find the other codimension-two saddle-node separatrix
loop point ($SNL_2$, at $S= -0.92$, not shown in Fig. \ref{fig:peix})
 where the SNIC line ends,
giving birth again to a homoclinic bifurcations line separated
from the saddle-node line. Following this line until S=-1 we
recover the $SL$ point at $Ra=1184$.

This particular configuration of codimension-two bifurcations is
very far from being peculiar of this problem. It is found in
many other areas, such as bursting oscillations in neural or
biological systems \citep{deVries,Borisuk}, population dynamics
\citep{Bazykin}, laser dynamics \citep{Mayol}, and some
diffusively coupled systems \citep{Kanamami}. Theoretically, all
this behavior can be reproduced as a 2-dimensional slice in
parameter space of the unfolding of a codimension-three degenerate
Takens-Bogdanov point \citep{DRS}, also known as the
Dumortier-Roussarie-Sotomayor (DRS) bifurcation. In particular, we
are referring to the focus case described by \citet{DRS}. This is
also consistent with the fact that the presence of the SNIC cannot
be related to a local codimension-2 phenomenon \citep{GJK2001}.

The scenario in which the DRS bifurcation appears can be described
by a planar vector field
\citep{DRS}, {\it i.e.} it is essentially two-dimensional.
This suggests the possibility of calculating the normal form
coefficients of the bifurcation from the PDE's, which is an
interesting problem from a theoretical point of view.

\section*{Acknowledgments}
This work was financially supported by
Direcci\'on General de
Investigaci\'on Cient\'{\i}fica y T\'ecnica (Spain)
(Projects BFM2003-00657 and  BFM2003-07850-C03-02) and
Comissionat per a Universitats i Recerca (Spain)
Projects (2001/SGR/00221 and 2002/XT/00010).
We also acknowledge computing support from
Centre Europeu de Paral·lelisme de Barcelona
(Spain).
E.M. acknowledges a grant from Ministerio de Educaci\'on, Cultura y Deporte
(Spain).

%
%
%
%
 \bibliographystyle{plainnat}
 \bibliography{mecatc}

\begin{thebibliography}{26}
\expandafter\ifx\csname natexlab\endcsname\relax\def\natexlab#1{#1}\fi
\expandafter\ifx\csname url\endcsname\relax
  \def\url#1{{\tt #1}}\fi

\bibitem[Bardan et~al.(2000)Bardan, Bergeon, Knobloch, and Mojtabi]{BaBeKnMo00}
G.~Bardan, A.~Bergeon, E.~Knobloch, and A.~Mojtabi.
\newblock Nonlinear doubly diffusive convection in vertical enclosures.
\newblock {\em Physica D}, 138:\penalty0 91--113, 2000.

\bibitem[Bazykin(1985)]{Bazykin}
A.~Bazykin.
\newblock {\em Mathematical Biophysics of interacting populations}.
\newblock Nauka, Moscow, 1985.

\bibitem[Bergeon et~al.(1998)Bergeon, Henry, BenHadid, and
  Tuckerman]{BeHeBeTu98}
A.~Bergeon, D.~Henry, H.~BenHadid, and L.S. Tuckerman.
\newblock Marangoni convection in binary mixtures with {Soret} effect.
\newblock {\em J. Fluid Mech.}, 375:\penalty0 143--177, 1998.

\bibitem[Bergeon and Knobloch(2002)]{BeKn02}
A.~Bergeon and E.~Knobloch.
\newblock Natural doubly diffusive convection in three-dimensional enclosures.
\newblock {\em Phys. Fluids}, 14:\penalty0 3233--3250, 2002.

\bibitem[Borisuk(1997)]{Borisuk}
Mark~T. Borisuk.
\newblock {\em Bifurcation Analysis of a Model of the Frog Egg Cell Cycle}.
\newblock PhD thesis, Virginia Polytechnic Institute and State University,
  1997.

\bibitem[Cross and Hohenberg(1993)]{cross93}
M.C. Cross and P.C. Hohenberg.
\newblock Pattern formation outside of equilibrium.
\newblock {\em Rev. Mod. Phys.}, 65\penalty0 (3):\penalty0 851--1112, 1993.

\bibitem[de~Vries(1996)]{deVries}
Gerda de~Vries.
\newblock {\em Analysis of the Models of Bursting Electrical Activity in
  Pancreatic $\beta$-Cells}.
\newblock PhD thesis, University of British Columbia, 1996.

\bibitem[Dumortier et~al.(1991)Dumortier, Roussarie, and Sotomayor]{DRS}
F.~Dumortier, R.~Roussarie, and J.~Sotomayor.
\newblock {\em Bifurcations of Planar Vector Fields}, chapter Generic
  3-Parameter Families of Planar Vector Fields, Unfoldings of Saddle, Focus and
  Elliptic Singularities With Nilpotent Linear Parts, pages 1--164.
\newblock Number 1480 in Lecture Notes in Mathematics. Springer-Verlag, 1991.

\bibitem[Frayss\'e et~al.(2003)Frayss\'e, Giraud, Gratton, and Langou]{cerfacs}
V.~Frayss\'e, L.~Giraud, S.~Gratton, and J.~Langou.
\newblock A set of {GMRES} routines for real and complex arithmetics on high
  performance computers.
\newblock Technical Report TR/PA/03/3, CERFACS, 2003.
\newblock Public domain software available on www.cerfacs/algor/Softs.

\bibitem[Ghorayeb and Mojtabi(1997)]{GhMo97}
K.~Ghorayeb and A.~Mojtabi.
\newblock Doubly diffusive convection in a vertical rectangular cavity.
\newblock {\em Phys. Fluids}, 9:\penalty0 2339--2348, 1997.

\bibitem[Golubistky et~al.(2001)Golubistky, Josic, and Kaper]{GJK2001}
M.~Golubistky, K.~Josic, and T.J. Kaper.
\newblock An unfolding theory aproach to bursting in slow-fast systems.
\newblock In H.~W. Broer, B.~Krauskopf, and G.~Vegta, editors, {\em Global
  Analysis of Dynamical Systems: Festschrift dedicated to Floris Takens on the
  occason of his 60th birthday}, pages 277--308. Institute of Physics
  Publications, 2001.

\bibitem[Hugues and Randriamampianina(1998)]{HuRa98}
S.~Hugues and A.~Randriamampianina.
\newblock An improved projection scheme applied to pseudospectral methods for
  the incompressible {Navier-Stokes} equations.
\newblock {\em Int. J. Numer. Methods Fluids}, 28:\penalty0 501--521, 1998.

\bibitem[Jiang et~al.(1991)Jiang, Ostrach, and Kamotani]{Jiang91}
H.~D. Jiang, S.~Ostrach, and Y.~Kamotani.
\newblock Unsteady thermosolutal transport phenomena due to opposed buoyancy
  forces in shallow enclosures.
\newblock {\em J. Heat Transfer}, 113:\penalty0 135, 1991.

\bibitem[Kanamami and Sekine(2003)]{Kanamami}
T.~Kanamami and M.~Sekine.
\newblock Array enhanced coherence resonance in the difussively coupled active
  rotators and its analysis with the nonlinear {Fokker}-{Plank} equation.
\newblock {\em IEICE Transactions on Fundamentals}, September 2003.

\bibitem[Knobloch et~al.(1986)Knobloch, Moore, Toomre, and Weiss]{KnMoToWe86}
E.~Knobloch, D.R. Moore, J.~Toomre, and N.O. Weiss.
\newblock Transition to chaos in two-dimensional double-diffusive convection.
\newblock {\em J. Fluid Mech.}, 166:\penalty0 409--448, 1986.

\bibitem[Kuznetsov(1998)]{Kuznetsov}
Y.A. Kuznetsov.
\newblock {\em Elements of Applied Bifurcation Theory}, volume 112 of {\em
  Applied Mathematical Sciences}.
\newblock Springer-Verlag, New York, 2 edition, 1998.

\bibitem[Mamun and Tuckerman(1995)]{MaTu95}
C.K. Mamun and L.S. Tuckerman.
\newblock Asymmetry and {Hopf} bifurcation in spherical {Couette} flow.
\newblock {\em Phys. Fluids}, 7:\penalty0 80--91, 1995.

\bibitem[Mayol et~al.(2002)Mayol, Toral, Mirasso, and Natiello]{Mayol}
C.~Mayol, R.~Toral, C.R. Mirasso, and M.A. Natiello.
\newblock Class a lasers with injected signal: Bifurcation set and {Lyapunov}
  potential function.
\newblock {\em Phys. Rev. A}, 66:\penalty0 013808 (1--12), 2002.

\bibitem[Meca et~al.(2004)Meca, Mercader, Batiste, and
  Ram\'{\i}rez-Piscina]{mecaprl}
E.~Meca, I.~Mercader, O.~Batiste, and L.~Ram\'{\i}rez-Piscina.
\newblock A blue sky catastrophe in double-diffusive convection.
\newblock {\em Physical Review Letters}, 2004.
\newblock (Submitted).

\bibitem[Schecter(1987)]{Schecter87}
S.~Schecter.
\newblock The saddle-node separatrix-loop bifurcation.
\newblock {\em SIAM J. Math. Anal.}, 18\penalty0 (4):\penalty0 1142--1156,
  1987.

\bibitem[Shilnikov(1997)]{LShil97}
L.~Shilnikov.
\newblock Mathematical problems of nonlinear dynamics: a tutorial.
\newblock {\em Int. J. Bifurcation and Chaos}, 7\penalty0 (9):\penalty0
  1953--2001, 1997.

\bibitem[Shilnikov and Turaev(2000)]{LShilTur00}
L.P. Shilnikov and D.V. Turaev.
\newblock A new simple bifurcation of a periodic orbit of "blue sky
  catastrophe" type.
\newblock {\em Amer. Math. Soc. Transl.}, 200\penalty0 (2):\penalty0 165--188,
  2000.

\bibitem[Turner(1980)]{Turner80}
J.S. Turner.
\newblock A fluid dynamical model of differential and layering in magna
  chambers.
\newblock {\em Nature}, 285:\penalty0 213--215, 1980.

\bibitem[Turner(1985)]{Turner85}
J.S. Turner.
\newblock Multicomponent convection.
\newblock {\em Ann. Rev. Fluid Mech.}, 17:\penalty0 11--44, 1985.

\bibitem[Xin and {Le Qu\'er\'e}(2001)]{XiQe01}
S.~Xin and P.~{Le Qu\'er\'e}.
\newblock Linear stability analyses of natural convection in a differentially
  heated square cavity with conducting horizontal walls.
\newblock {\em Phys. Fluids}, 13:\penalty0 2529--2542, 2001.

\bibitem[Xin et~al.(1997)Xin, {Le Qu\'er\'e}, and Tuckerman]{XiQeTu97}
S.~Xin, P.~{Le Qu\'er\'e}, and L.~Tuckerman.
\newblock Bifurcation analysis of doubly-diffusive convection with opposing
  horizontal thermal and solutal gradients.
\newblock {\em Phys. Fluids}, 10:\penalty0 850--858, 1997.

\end{thebibliography}
%
%
%

\end{document}